\documentclass[aps,preprint,groupedaddress]{revtex4}

\usepackage{graphicx}
\usepackage[fleqn]{amsmath}
\usepackage{amsfonts}
\usepackage{amssymb}
\usepackage{psfrag}

\newcommand{\Diff}[2]{\frac{\partial #1}{\partial #2}}

\newcommand{\At}[1]{\biggr\rvert_{#1}}
\DeclareMathOperator{\Erf}{Erf}

\newcommand{\Ds}{\displaystyle}

\newcommand{\Dip}{_{\text{dip}}}
\newcommand{\Sr}{_{\text{sr}}}

\newcommand{\Tens}[1]{\boldsymbol{\rm #1}}
\newcommand{\Vect}[1]{\boldsymbol{\rm #1}}
\newcommand{\Ver}{\Vect{r}}
\newcommand{\Vmu}{\Vect{\mu}}
\newcommand{\Vp}{\Vect{P}}
\newcommand{\Grad}{\Vect{\nabla}}
\newcommand{\Rom}[1]{\mathrm{#1}}
\newcommand{\Romd}{\Rom{d}}
\newcommand{\Lapl}{\Delta}

\begin{document}



\title{Onsager model for a variable dielectric permittivity near an interface}

\author{Reimar Finken}
\email{rf227@cam.ac.uk}
\author{Vincent Ballenegger}
\author{Jean-Pierre Hansen}

\affiliation{Department of Chemistry\\
University of Cambridge\\
Cambridge CB2 1EW (UK)}

\date{\today}

\begin{abstract}
Using a generalisation of an Onsager type approach, we are able to predict a dielectric
permittivity profile of an inhomogeneous dipolar fluid in the presence
of a dielectric interface. The reaction and cavity fields are
calculated semi-analytically using bispherical coordinates. An
asymptotic expression for the local permittivity is derived as a
function of distance from the interface.
\end{abstract}

\maketitle


\section{Introduction}
\label{sec:introduction}

Since Onsager \cite{onsager36} and Kirkwood \cite{kirkwood39} it is well
known that the macroscopic dielectric constant (or permittivity)
$\epsilon$ of a material of polar, polarizable molecules is related to
the fluctuations of the total dipole moment of a spatially homogeneous
sample \cite{madden84}. While the relation is well understood for infinite
or periodic \cite{neumann83} bulk materials, much less is known about the
static dielectric properties of confined polar fluids, or polar fluids
near interfaces. The generally accepted view is that near interfaces,
the dipole moment fluctuations are restricted, e.g. when dipole
moments of individual molecules tend to orient themselves
preferentially parallel or perpendicular to an interface, and that the
resulting local permittivity is reduced relative to its bulk value.
This is particularly important when considering water molecules near
electrodes, biological macromolecules (e.g.
proteins or DNA) or membranes in electrochemical and biophysical
applications. On purely phenomenological grounds it is often assumed
that the dielectric response of a polar solvent in such situations may
be described by a distance-dependent permittivity, which is then
incorporated in expressions for the Coulomb interaction energy between
ions in solution \cite{schaefer99}.

The present paper attempts a first step beyond such purely {\em ad
hoc} procedures, in terms of a simplified molecular picture of a
polar fluid near an interface, by generalising Onsager's well-known
cavity model for bulk dielectric properties to an inhomogeneous
situation. The objective is to determine a permittivity ``profile''
which varies with the distance from a planar interface separating a
continuous medium of given permittivity $\epsilon_1$ from the medium
made up of dipolar molecules.

The paper is organised as follows: The generalised Onsager cavity
model is described in section \ref{sec:onsager-model-near}. A
self-consistent set of equations determining the permittivity profile
is derived in section \ref{sec:reaction-field}. In section
\ref{sec:asympt-behav-diel} the asymptotic behaviour of the dielectric
permittivity far away from the interface is examined. The solution to
the full electrostatic problem leading to a locally self-consistent
permittivity profile is sketched in section
\ref{sec:solut-electr-probl}. The results will be discussed in the
final section.

\section{Onsager model near a planar interface}
\label{sec:onsager-model-near}

\begin{figure}[htbp]
  \centering
  \psfrag{theta}{\mbox{{\Large $\vartheta$}}}
  \psfrag{m}{\mbox{{\Large $\Vect{\mu}$}}}
  \psfrag{z}{\mbox{{\Large $z$}}}
  \psfrag{R}{\mbox{{\Large $R$}}}
  \psfrag{E}{\mbox{{\Large $E$}}}
  \psfrag{e}{\mbox{{\Large $\epsilon$}}}
  \includegraphics[width=\textwidth]{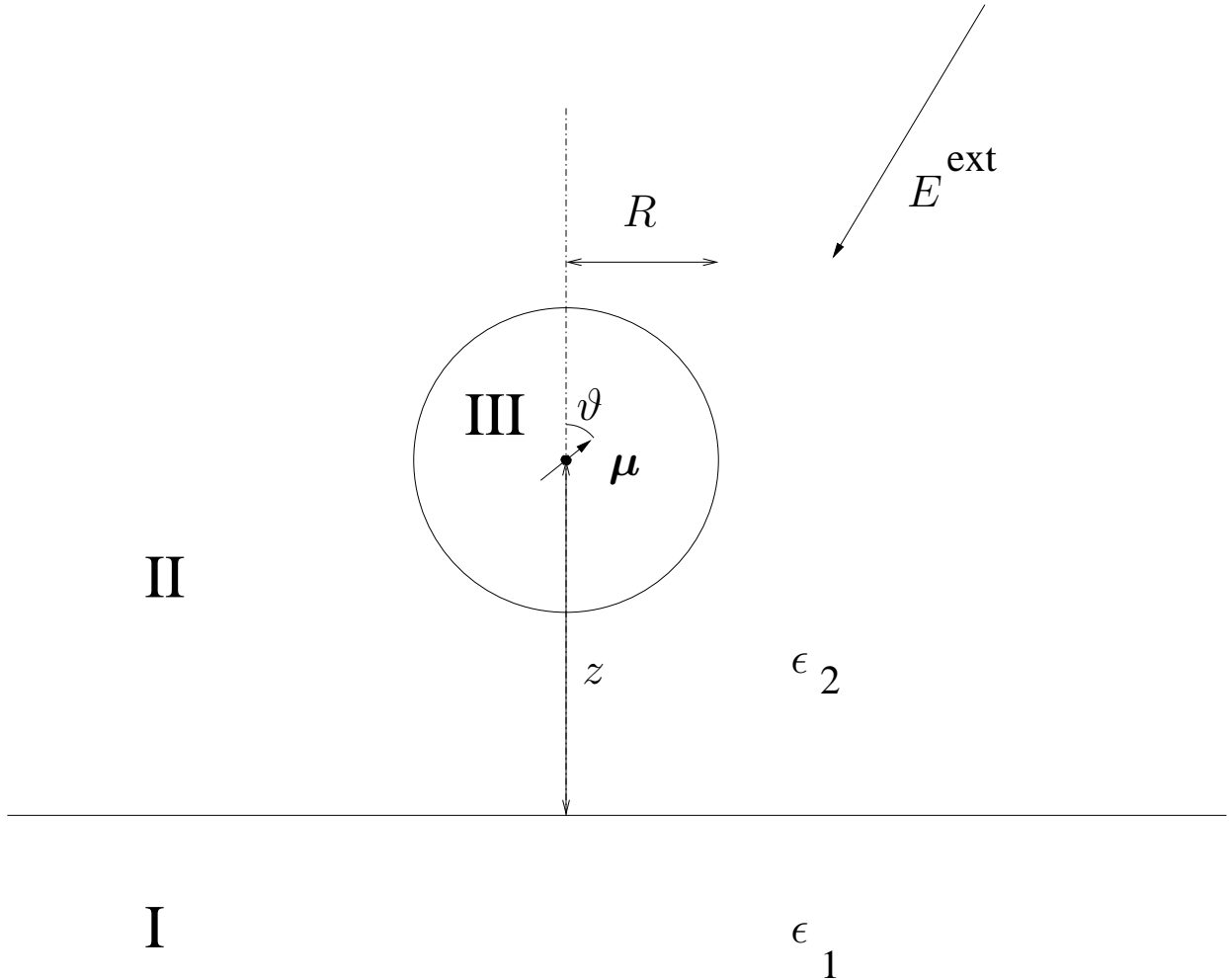}
  \caption{Geometry of the Onsager model. The test dipole $\Vect{\mu}$
  is in a spherical cavity with centre a distance $z$ above the
  dielectric interface, below which extends a continuum of
  permittivity $\epsilon_1$. The fluid in region II has a local
  permittivity tensor $\Tens{\epsilon}_2$. Far above the wall the
  electric field is the external field $\Vect{E}^{\text{ext}}$. The
  polar angle of dipole with respect to the $z$ axis is denoted by $\vartheta$.}
  \label{fig:1}
\end{figure}

Consider two dielectric media separated by a planar interface, as
shown in figure \ref{fig:1}. The horizontal interface is placed at
$z=0$. Below it ($z < 0$) extends a dielectric medium considered to be
a homogeneous continuum of constant permittivity $\epsilon_1$. Above
the interface ($z>0$) extends a polar fluid made up of identical
spherical molecules carrying a dipole moment $\Vect{\mu}$ of fixed
magnitude $\mu = |\Vect{\mu}|$. The interface breaks the rotational
invariance of the fluid, so that its dielectric response is {\em a priori}
described by a permittivity tensor $\Tens{\epsilon}_2$ which, by
symmetry, is diagonal and of the form
\begin{equation}
  \label{eq:1}
  \Tens{\epsilon}_2 =  \begin{pmatrix}
    \epsilon^{\|}_2 & 0 &0\\
    0 & \epsilon^{\|}_2&0\\
    0 & 0 & \epsilon^{\bot}_2
  \end{pmatrix},
\end{equation}
where $\epsilon^{\|}_2$ denotes the identical $xx$ and $yy$ components
parallel to the interface, while $\epsilon^{\bot}_2$ denotes the
vertical $zz$ component.

In view of an approximate calculation of $\epsilon^{\|}_2$ and
$\epsilon^{\bot}_2$, one assumes, following Onsager \cite{onsager36},
that a test molecule is placed at the centre of a spherical cavity of
radius $R = (3v/4\pi)^{1/3}$, where $v=V/N$ is the volume per molecule. Since we are interested in permittivity
profiles, the centre of the cavity is placed at a vertical distance
$z$ from the planar interface, and the dielectric response will be
determined as a function of $z$. The cavity is surrounded by a
dielectric continuum characterised by the dielectric tensor
$\Tens{\epsilon}_2$, of yet unknown components $\epsilon^{\|}_2$ and
$\epsilon^{\bot}_2$. Following Onsager's mean field approach
\cite{onsager36}, the dipole $\Vect{\mu}$ inside the cavity is
subjected to a cavity field $\Vect{E}_C$, induced by some external
field taking the value $\Vect{E}^{\text{ext}}$ far above the planar
interface (i.e. at $z \rightarrow +\infty$). In addition, the dipole polarises
the surrounding dielectric media I and II, giving rise to polarisation
charges on the surface of the cavity and on the planar interface.
These polarisation charges generate a reaction field $\Vect{E}_R$
acting on the dipole inside the cavity. Both fields $\Vect{E}_C$ and
$\Vect{E}_R$ are calculated from the laws of macroscopic
electrostatics. The mean dipole moment inside the cavity is then
calculated by averaging over orientations with the appropriate
Boltzmann weight determined by the coupling of the dipole to the total
electric field $\Vect{E} = \Vect{E}_C + \Vect{E}_R$ inside the cavity.
The dielectric tensor components $\epsilon^{\|}_2$ and
$\epsilon^{\bot}_2$ are finally determined by identifying the
calculated mean dipole with the predictions of macroscopic
electrostatics for parallel and orthogonal orientations of the
external field.

The calculation sketched above is quite straightforward in the case of
isotropic, bulk dielectrics, but, as will become clear below, it is
technically much more involved in the presence of an interface
between two different dielectric media. 

In the isotropic bulk, where $\epsilon^{\|}_2 = \epsilon^{\bot}_2 =
\epsilon_2$, the Onsager mean field theory leads to the classic result
\cite{onsager36} 
\begin{equation}
  \label{eq:2}
  \frac{(\epsilon_2 -1)(2\epsilon_2 +1)}{\epsilon_2} = \frac{\rho
  \mu^2}{\epsilon_0 k_B T},
\end{equation}
valid for non-polarizable, polar molecules, $\rho=1/v$ being the number
density of these molecules and $k_B T$ the thermal energy; $\epsilon_0$
is the vacuum permittivity. The predictions of \eqref{eq:2}, which
does not account for correlations between the orientations of
neighbouring dipoles (as embodied in the Kirkwood $g_K$-factor, see eq. \eqref{Kirk}), considerably underestimate the value of
$\epsilon_2$ for dense, highly polar liquids, and the same is hence to
be expected for the results near the planar interface.

\section{Self-consistent equation for the permittivity profile}
\label{sec:reaction-field}

In the presence of the planar interface (cf. Fig. \ref{fig:1}), the polarisation vector $\Vect{P}$ in region~II depends on the vertical coordinate $z$. If $\Vect{E}(z)$ is the local electric field, the local susceptibility and dielectric tensors $\Tens{\chi}_2$ and $\Tens{\epsilon}_2$ can be defined as usual by the linear relation
\begin{equation}
  \label{eq:3}
  \Vect{P}(z) = \epsilon_0 \Tens{\chi}_2(z) \cdot \Vect{E}(z) =
  \epsilon_0 (\Tens{\epsilon}_2(z) - \Tens{I}) \cdot \Vect{E}(z),
\end{equation}
assuming a purely local response.

As explained in the previous section, the local field acting on the dipole $\Vect{\mu}$ inside the cavity is the sum of the cavity and reaction fields.
Due to the linear nature of electrostatics, the reaction field depends linearly on the dipole moment
\begin{equation}
  \label{eq:4}
  \Vect{E}_R = \Tens{M}_R \cdot \Vect{\mu}.
\end{equation}
The tensor $\Tens{M}_R$ depends on $\epsilon_1$ and $\Tens{\epsilon}_2$ and on the geometric parameters $R$ and $z$. Due to the cylindrical symmetry of the
problem, it is clear that $\Tens{M}_R$ must be diagonal. A vertical
dipole (parallel to the symmetry axis $Oz$) can only induce a vertical
reaction field. For an arbitrary orientation of $\Vect{\mu}$, the
reaction field must change sign under the inversion of the dipole
$\Vect{\mu} \rightarrow -\Vect{\mu}$. In the case of a horizontal
dipole, this inversion is equivalent to a rotation by $\pi$ around the
vertical symmetry axis, implying that the reaction field has no vertical component. The horizontal component must moreover be parallel to the dipole, since the system is invariant under reflections by the plane containing $\Vect{\mu}$ and the vertical symmetry axis. The tensor
$\Tens{M}_R$ is hence of the same diagonal form \eqref{eq:1} as the
permittivity tensor, with one vertical component $M_R^{\bot} =
M_{Rzz}$, and two horizontal components $M_R^{\|}=M_{Rxx}=M_{Ryy}$.

Similarly, the cavity field is proportional to the external field
\begin{equation}
  \Vect{E}_C = \Tens{M}_C \cdot \Vect{E}^{\text{ext}},
\end{equation}
where the tensor $\Tens{M}_C$ can only depend on $\epsilon_1$,
$\epsilon_2$, and on the geometric parameters $R$ and $z$. The same symmetry arguments as adove show that the tensor $\Tens{M}_C$ is diagonal.

If the dipole interact with the cavity field via the energy $- \Vect{\mu} \cdot \Vect{E}_C$, the interaction energy with the reaction field is given by $- \Vect{\mu} \cdot \Vect{E}_R/2$. Indeed, increasing gradually the dipole moment from zero to its final value, we obtain a factor $1/2$ when we integrate the infinitesimal interaction energy increment $-\Vect{E}_R \cdot d\Vect{\mu}$ because of the linear relation \eqref{eq:4}.

The induced dipole moment per unit volume (or polarisation) is, within
the cavity model,
\begin{equation}
  \label{eq:5}
  \Vect{P} = \rho \langle\Vect{\mu}\rangle = \rho \frac{\int
  \Vect{\mu} \exp\{\beta \Vect{E}_C\cdot \Vect{\mu}\} \exp\{\beta
  \Vect{E}_R \cdot \Vect{\mu}/2\} d\Omega}{\int
  \exp\{\beta \Vect{E}_C\cdot \Vect{\mu}\} \exp\{\beta \Vect{E}_R
  \cdot \Vect{\mu}/2\}d\Omega},
\end{equation}
where the integrations are over all orientations of the test dipole.
For small external field, the polarisation will be proportional to the
field, and we can expand the exponential to first order in the reaction field. Taking into account the diagonal nature of the cavity field tensor
$\Tens{M}_C$, we obtain in linear response
\begin{align}
  \Vect{P} &= \rho \frac{\int \Vect{\mu} \exp\{\beta \Vect{E}_R \cdot
    \Vect{\mu}/2\} (1+\beta \Vect{E}_C\cdot \Vect{\mu})d\Omega}{\int \exp\{\beta \Vect{E}_R \cdot
    \Vect{\mu}/2\} (1+\beta \Vect{E}_C\cdot \Vect{\mu})d\Omega}\nonumber\\
  &= \rho \frac{\int \Vect{\mu} \exp\{\beta \Vect{E}_R \cdot
    \Vect{\mu}/2\} \beta \Vect{E}_C\cdot \Vect{\mu}d\Omega}{\int \exp\{\beta \Vect{E}_R \cdot
    \Vect{\mu}/2\}d\Omega}\nonumber\\
  &= \beta \rho \frac{\int \Vect{\mu} \Vect{\mu}
    \exp\{\beta\Vect{\mu}\cdot\Tens{M}_R\cdot\Vect{\mu}/2\} d\Omega}{\int 
    \exp\{\beta\Vect{\mu}\cdot\Tens{M}_R\cdot\Vect{\mu}/2\}
    d\Omega}\cdot \Tens{M}_C \cdot \Vect{E}^{\text{ext}}\nonumber\\
  &= \epsilon_0 \,\Tens{\chi}_2^{\text{ext}} \cdot \Vect{E}^{\text{ext}}.  \label{eq:6}
\end{align}
In going from the first to the second line of \eqref{eq:6}, integrals involving an odd
function of $\Vect{\mu}$  vanish. In the last expression we have defined the
dielectric susceptibility with respect to the external field
\begin{equation}
\label{chi_2^ext}
  \Tens{\chi}_2^{\text{ext}} = \frac{\beta \rho \mu^2}{\epsilon_0} \frac{\int \hat{\Vect{\mu}} \hat{\Vect{\mu}}
    \exp\{\beta\Vect{\mu}\cdot\Tens{M}_R\cdot\Vect{\mu}/2\} d\Omega}{\int 
    \exp\{\beta\Vect{\mu}\cdot\Tens{M}_R\cdot\Vect{\mu}/2\}
    d\Omega}\cdot \Tens{M}_C.
\end{equation}
The components of this tensor are
\begin{equation}
\label{comp chi_2^ext}
   \chi_{\|}^{\text{ext}} = \frac{9y}{2} (1-\Delta_R) M_C^{\|},
   \qquad
   \chi_{\bot}^{\text{ext}} = \frac{9y}{2} \Delta_R M_C^{\bot},
\end{equation}
with $y=\beta \rho \mu^2/\epsilon_0$ and
\begin{equation}
\Delta_R = \frac{1}{2\alpha^2} -
\frac{e^{-\alpha^2}}{\sqrt{\pi}\alpha\Erf(\alpha)},\quad \alpha=\sqrt{\mu^2(M_R^{\|}-M_R^{\bot})/2}.
  \label{eq:7}
\end{equation}
($\Erf$ denotes the error function). The external susceptibility \eqref{chi_2^ext} must now be related to the local susceptibility $\Tens{\chi}_2$ defined in \eqref{eq:3}.

In the present case of a planar interface, it is easy to find the macroscopic relation between the polarisation $\Vect{P}(z)$ and the external field. For a given field far away from the wall and a given permittivity profile, the electric field at every point is known from the macroscopic equations. When the external field is horizontal, it will stay
constant throughout space, since the permittivity is only varying
perpendicularly to the field. Therefore the field felt by the test
dipole in the macroscopic picture is simply $\Vect{E}^{\text{ext}}$,
leading to the polarisation $\Vect{P} = \epsilon_0
(\epsilon_2^{\|}(z)-1) \Vect{E}^{\text{ext}}$. In the case of a
vertical field, however, the displacement field
\begin{equation}
   \Vect{D}(z)\equiv\epsilon_0 \Vect{E}(z) + \Vect{P}(z) = \epsilon_0 \, \Tens{\epsilon}_2 \cdot \Vect{E}(z)
\end{equation} will remain
constant. Therefore, at distance $z$ from the interface, the electric field is
$\Vect{E}^{\text{ext}}\,\epsilon_2^{\bot}(\infty)/\epsilon^{\bot}_2(z)$,
leading to the polarisation $\Vect{P}= \epsilon_0
(\epsilon^{\bot}_2(\infty)-\frac{\epsilon_2^{\bot}(\infty)}{\epsilon_2^{\bot}(z)})\Vect{E}^{\text{ext}}$.
The permittivity far away from the wall is simply the bulk
permittivity.

Comparing the above expression of the polarisation with the result \eqref{comp chi_2^ext} obtained from Onsager's model gives
\begin{subequations}
\begin{align}
  \label{eq:8}
  \chi^{\text{ext}}_{2,\|} &= \epsilon_2^{\|}(z) -1 =  \frac{9y}{2} (1-\Delta_R) M_C^{\|} \\
  \chi^{\text{ext}}_{2,\bot} &= \epsilon_2^{\bot}(\infty)
  -\frac{\epsilon_2^{\bot}(\infty)}{\epsilon_2^{\bot}(z)} = \frac{9y}{2} \Delta_R M_C^{\bot}.
  \label{eq:9}
\end{align}
\end{subequations}
The tensors $\Tens{M}_C$ and $\Tens{M}_R$ [which determine $\Delta_R$ via \eqref{eq:7}] are to be obtained by solving the associated electrostatic problems. Since the permittivity tensor $\Tens{\epsilon}_2(z)$ itself enters these problems, equations \eqref{eq:8} and \eqref{eq:9} are self-consistent equations for this permittivity profile.

Before determining the cavity and reaction field tensors in section~V, we obtain first the asymptotic behaviour of the permittivity. 

\section{Asymptotic behaviour of the dielectric permittivity}
\label{sec:asympt-behav-diel}

It is instructive to obtain an approximate estimate
of the corrections to the reaction and cavity fields, valid for a test dipole far away from the
interface, such that $z \gg R$. The dielectric tensor will then
practically reduce to the bulk value $\epsilon_2 \Tens{I}$. The
polarisation of the cavity surface by the test dipole in the absence
of an external field is such that the
latter acts outside the cavity as an effective dipole moment of strength
$\frac{3}{2\epsilon_2 + 1}\mu$. The cavity model
only yields an approximate estimate of
 the effective dipole moment. An
exact calculation, sketched in Appendix \ref{app:a}, yields a general
expression for the screened (effective) dipole moment which reduces to
the prediction of the cavity model if the Onsager value \eqref{eq:2}
for $\epsilon_2$ is substituted in equation \eqref{p^eff}.

\begin{figure}[htbp]
  \centering
  \psfrag{theta}{\LARGE $\vartheta$}
  \psfrag{ereact}{\LARGE $-\Vect{\mu}\cdot \Vect{E}_R / (2 k_B T)$}
  \includegraphics[angle=-90,width=\textwidth]{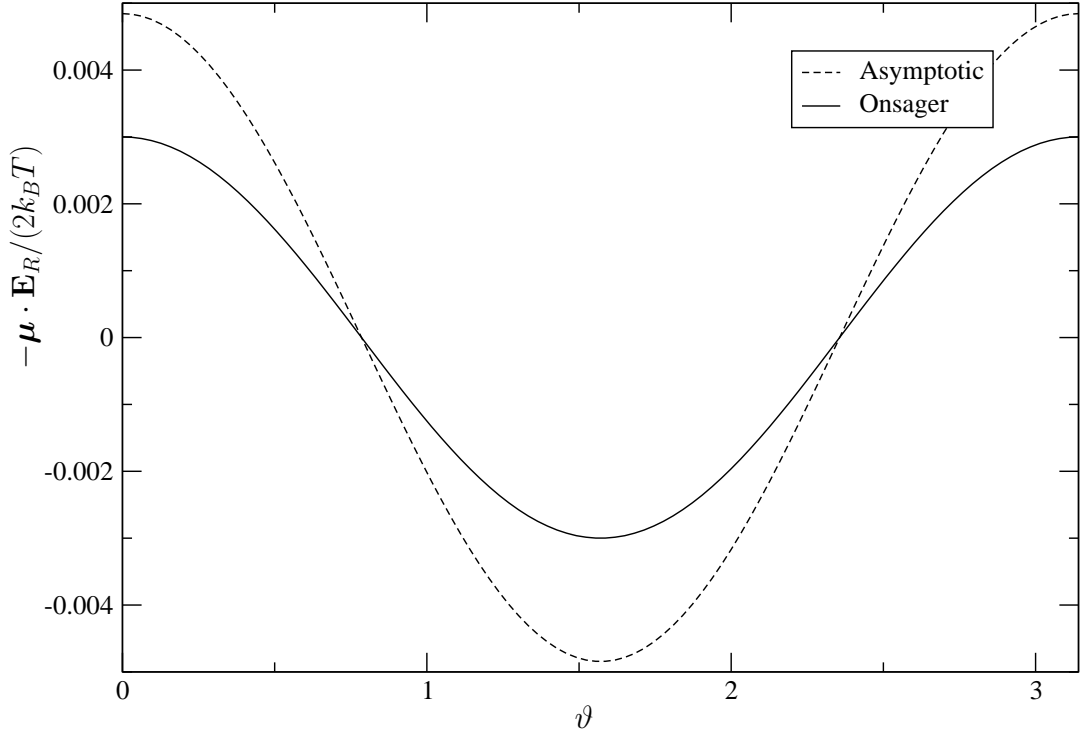}
  \caption{Potential energy $-\Vect{\mu} \cdot \Vect{E}_R /2$ divided
  by the thermal energy $k_B T$ as a function of the azimuthal angle
    $\vartheta$ due to the interaction of the dipole with the reaction
    field as predicted by the Onsager model (solid line) and as given
    by the asymptotic expansion \eqref{eq:10} (dashed line). The dipole is located in a distance $2R$ above the wall.
    Dipole moment $\mu=1.86\mathrm{D}=6.2\cdot10^{-30} \mathrm{Cm}$ and number density
    $\rho=3.346\cdot 10^{28} \mathrm{m}^{-3}$ are the experimental values
    measured for water at $T=293\mathrm{K}$. This leads in the bulk to an
    Onsager value of $\epsilon_2=18.4$. The material below the
    interface is vacuum $\epsilon_1=1$. The constant
    electrostatic interaction energy which is also present in the
    bulk has been subtracted. The amplitude of the orientation dependent energy is
    less than $0.01k_B T$ even in the immediate vicinity of the wall.}
  \label{fig:2}
\end{figure}

The additional electric field generated by the polarisation of the planar
interface at $z=0$ by this effective dipole may be described as being
due to an image dipole of strength $\frac{\epsilon_2 -
  \epsilon_1}{\epsilon_2+\epsilon_1}\frac{3}{2\epsilon_2 + 1}\mu$
situated at $-z$ below the interface. Since $z/R \gg 1$, the resulting
field may be regarded as homogeneous over a distance equal to the
diameter $2R$ of the cavity. The reaction field inside the cavity will
be that outside the cavity, amplified by a factor $3
\epsilon_2/(2\epsilon_2+1)$. The field depends of course on the
orientation of the image dipole which itself is determined by the
orientation of the test dipole. If the
test dipole has both vertical and horizontal components,
$\Vect{\mu}^{\bot}$ and $\Vect{\mu}^{\|}$, the reaction field is therefore
\begin{equation}
  \label{eq:10}
  \Vect{E}_R = \frac{1}{4\pi\epsilon_0}\biggl[2
    \frac{\epsilon_2-1}{2\epsilon_2 + 1} \frac{\Vect{\mu}}{R^3}
    -\left(\frac{3}{2\epsilon_2+1}\right)^2 \epsilon_2
    \frac{\epsilon_2-\epsilon_1}{\epsilon_2+\epsilon_1}
    \frac{1}{8 z^3} \bigl[\Vect{\mu}^{\|} + 2 \Vect{\mu}^{\bot}\bigr]\biggr],
\end{equation}
and hence is no longer collinear with $\Vect{\mu}$. The electrostatic
interaction energy of the dipole with the reaction field
$-\Vect{\mu}\cdot \Vect{E}_R$ therefore depends on the orientation of
the molecule. In figure \ref{fig:2} it can be seen that the
orientation dependence of the energy is negligible even in the
immediate vicinity of the wall.

In the presence of an external field two parts are contributing to the
effective dipole moment. The test molecule aligns on average with the
cavity field, leading to a permanent dipole, which appears screened
outside the cavity. A second contribution originates from the
polarisation of the cavity surface by the external field. Both
terms lead to an electric field outside the cavity as due to an effective dipole
moment
\begin{equation}
  \label{eq:11}
  \Vect{\mu}^{\text{eff}} = \left[\frac{3}{2\epsilon_2+1} \frac{\mu^2}{3 k_B
  T} \frac{3 \epsilon_2}{2 \epsilon_2 + 1} - 4 \pi \epsilon_0
\frac{\epsilon_2-1}{2\epsilon_2+1}R^3\right]\Vect{E}^{\text{ext}}.
\end{equation}
The corresponding image dipole $\mu^{\text{img}} =
\frac{\epsilon_2-\epsilon_1}{\epsilon_2+\epsilon_1}\mu^{\text{eff}}$
induces an approximately homogeneous field $\Vect{E}^{\text{img}}$
near the cavity, which is the first order correction of the local
electric field $\Vect{E}^{\text{loc}} =
\Vect{E}^{\text{img}}+\Vect{E}^{\text{ext}}$ over the external field.
From the local field the local polarization can be determined to first
order in $z^{-3}$
\begin{align}
  \label{eq:12}
  \Vect{P} &= \rho \langle\Vect{\mu}\rangle =  \frac{\rho \beta \mu^2
    \epsilon_2}{2\epsilon_2+1} \Vect{E}^{\text{loc}}\\
  &= 9\epsilon_0 y \frac{\epsilon_2}{2\epsilon_2+1} \left\{
  \Vect{E}^{\text{ext}} - \frac{\Vect{E}^{\text{ext}}_{\|}+2\Vect{E}^{\text{ext}}_{\bot}}{32 \pi \epsilon_0
  z^3}\frac{\epsilon_2-\epsilon_1}{\epsilon_2+\epsilon_1}
  \left[\frac{3 \epsilon_2 \beta
  \mu^2}{(2\epsilon_2+1)^2}-4\pi\epsilon_0 R^3 \frac{\epsilon_2-1}{2\epsilon_2+1}\right]\right\}.
\end{align}
After extracting $\Tens{\chi}_2^{\text{ext}}$ we can find the dielectric
permittivities to first order in $z^{-3}$ as
\begin{align}
  \label{eq:13}
  \epsilon_2^{\|}(z) - \epsilon_2(\infty) &=  - \frac{\epsilon_2 - 1}{32 \pi \epsilon_0
    z^3}\frac{\epsilon_2-\epsilon_1}{\epsilon_2+\epsilon_1}
  \left[\frac{3 \epsilon_2 \beta
      \mu^2}{(2\epsilon_2+1)^2}-4\pi\epsilon_0 R^3
    \frac{\epsilon_2-1}{2\epsilon_2+1}\right]\\
   \epsilon_2^{\bot}(z)- \epsilon_2(\infty) &= - \frac{\epsilon_2
    (\epsilon_2 -1)}{16 \pi \epsilon_0
  z^3}\frac{\epsilon_2-\epsilon_1}{\epsilon_2+\epsilon_1}
  \left[\frac{3 \epsilon_2 \beta
  \mu^2}{(2\epsilon_2+1)^2}-4\pi\epsilon_0 R^3 \frac{\epsilon_2-1}{2\epsilon_2+1}\right].
\end{align}
All quantities appearing on the r.h.s in these expressions are bulk
values.

\section{Solution of the electrostatic problem}
\label{sec:solut-electr-probl}

To find the tensors $\Tens{M}_C$ and $\Tens{M}_R$ we have to solve two purely
electrostatic problems. One involves finding the reaction field
induced by the polarised material at the centre of the cavity.
In this case we consider a dipole at the centre of the spherical cavity,
surrounded by dielectric material of permittivity $\Tens{\epsilon}_2$
at a distance $z$ above a medium with isotropic permittivity
$\epsilon_1$. The electric field must decay to
zero as $z \rightarrow \infty$. To find the cavity field on the other hand, we consider an empty
sphere, and an external field which goes to $\Vect{E}^{\text{ext}}$
far from the cavity. In both problems the
electrostatic equations are the same; they differ only in the boundary
conditions: for the cavity field we require the electric field to be
finite within the sphere and approach a prescribed value at infinity;
for the reaction field we require a dipolar singularity at the centre
of the sphere and the electric field vanishes far away from the interface. 

In the geometry considered here there are two dielectric interfaces: One is
a planar interface where the dielectric permittivity jumps from
$\epsilon_1$ to $\Tens{\epsilon}_2$. The other is a sphere where the
permittivity jumps from $\Tens{\epsilon}_2$ to 1. These interfaces
divide space into three regions 
as shown in figure \ref{fig:1}. Within each region the electrostatic
potential satisfies the relation
\begin{equation}
\Grad\cdot(\Tens{\epsilon}\cdot\Grad \psi)=0.\label{eq:17}
\end{equation}
With a spatially varying tensor permittivity $\Tens{\epsilon}_2(z)$, eq. \eqref{eq:17} cannot be solved analytically. Instead of computing a numerical solution, we shall determine the cavity and reaction fields for the simpler case of a \emph{constant} and \emph{scalar} permittivity $\epsilon_2$. These results will then be used to determine a permittivity profile from the Onsager equations \eqref{eq:8} and \eqref{eq:9}. These equations will be solved in a locally self-consistent manner: when the cavity is at a distance $z$ from the interface, we shall use the cavity and reaction field tensors corresponding to the case of a homogeneous and isotropic medium II of dielectric constant $\overline{\epsilon_2}\equiv\text{Tr}(\Tens{\epsilon}_2(z))/3 = [2
\epsilon_2^{\|}(z)+\epsilon_2^{\bot}(z)]/3$. This scheme provides a reasonnable approximation to $\Tens{\epsilon}(z)$, since the deviations of $\Tens{\epsilon}(z)$ from the bulk behaviour $\epsilon_2 \Vect{I}$ prove to be small for $z>2R$ (see section~VI). 

In each regions, equation \eqref{eq:17} now reduces to Laplace's equation
\begin{equation}
\Delta \psi = 0 \label{eq:18}.
\end{equation}
At the dielectric interfaces we impose the usual
boundary conditions: The tangential component of the electric field
$\Vect{E} = -\Grad \psi$ must be identical on both sides of the
interface:
\begin{equation}
  \label{eq:19}
  \Grad \psi\At{t,I} = \Grad\psi\At{t,II}
\end{equation}
This condition is automatically satisfied if the potential is continuous across the interface:
\begin{equation}
  \label{eq:20}
  \psi\At{I} = \psi\At{II}
\end{equation}
Likewise the normal part of the displacement field
$\Vect{D} = {\epsilon}_0{\epsilon}_2 \Vect{E}$ must be identical on both
sides of the boundary. 
\begin{equation}
  \label{eq:21}
   {\epsilon}_1\, \Grad \psi\At{n,I} = {\epsilon}_{2} \,\Grad \psi\At{n,II}
\end{equation}

\begin{figure}[htbp]
  \centering
  \psfrag{a}{\mbox{{\Large $a$}}}
  \psfrag{e}{\mbox{{\Large $\eta$}}}
  \includegraphics[width=0.8\textwidth]{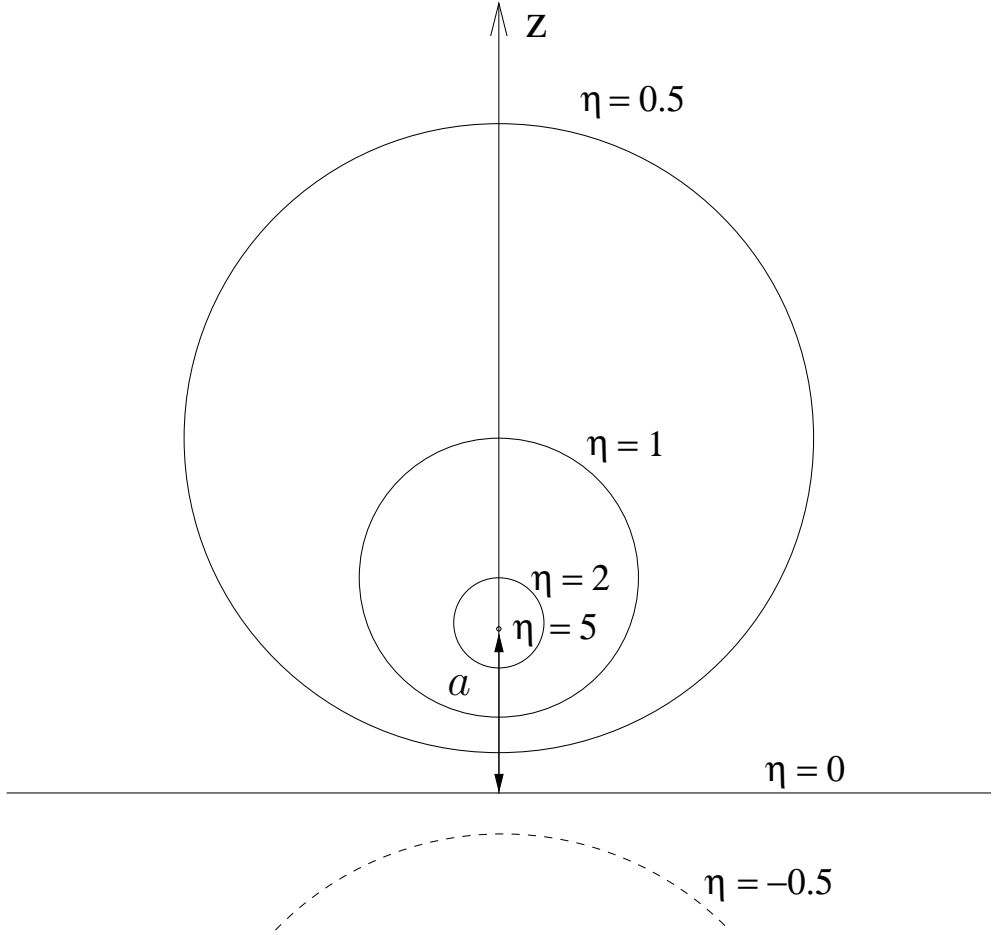}
  \caption{Surfaces of constant $\eta$ in bispherical coordinates.
  A cut through the cartesian $xz$-plane is shown. In three dimensions
  the surfaces are spheres. As $\eta\rightarrow \infty$, the centres
  of the spheres approach the point $z=a$.}
  \label{fig:3}
\end{figure}

The electrostatic problem is best solved in bispherical coordinates. Some properties of this system are summarised in Appendix \ref{app:b}. This coordinate system is well suited
for our problem because the two interfaces turn out to be surfaces of
constant coordinate $\eta$. A set of these surfaces is sketched in
figure \ref{fig:3}. Furthermore Laplace's
equation is separable in these coordinates. We therefore proceed to express the
yet unknown electrostatic potential in each region as a series of
fundamental solutions of Laplace's equation with unknown expansion
coefficients. The latter will be determined by the boundary conditions. In bispherical coordinates the component of the displacement field normal to the surface of constant coordinate $\eta$ is given by~\footnote{For a tensor permittivity $\Tens{\epsilon}_2$, the factor $\epsilon_2$ in this equation would be replaced by $\epsilon_2^{\|}$, and there would be an additional term
$
  - \frac{\epsilon_{\bot} - \epsilon_{\|}}{a} \frac{1-\cosh \eta \cos \vartheta}{\cosh \eta - \cos \vartheta} \left[(1-\cosh \eta \cos
    \vartheta) \Diff{\psi}{\eta} + \sinh \eta \sin \theta
    \Diff{\psi}{\vartheta}\right]
$.
}
\begin{equation}
  \label{eq:22}
  D_\eta = - \frac{\cosh \eta - \cos \vartheta}{a} \epsilon_0\epsilon_2
  \Diff{\psi}{\eta}.
\end{equation}

We fix the parameter $a$ in the bispherical coordinates to $a =
R\sqrt{z^2/R^2-1}$ and define $\eta_0 = \ln(z/R+\sqrt{(z/R)^2-1})$. With
these choices the planar interface corresponds to the surface
$\eta=0$, and the spherical cavity to the surface $\eta =
\eta_0$. The centre of the sphere has coordinates $\eta_1 = 2\eta_0$,
$\vartheta_1=0$, and $\phi_1=0$. The solutions of Laplace's
equation in bispherical coordinates are of the form
\begin{equation}
  \label{eq:23}
  \psi(\eta,\vartheta,\phi) = \sum_{n=0}^\infty \sum_{m=-n}^n \left[A_{mn} \psi_{mn}^{-}(\eta,\vartheta,\phi) +
  B_{mn} \psi_{mn}^{+}(\eta,\vartheta,\phi)\right],
\end{equation}
with the fundamental solutions 
\begin{equation}
  \label{eq:24}
  \psi^{\pm}_{mn}(\eta, \vartheta, \phi) = \sqrt{\cosh \eta - \cos \vartheta} e^{\pm (n +
    1/2) \eta} Y^m_n(\vartheta, \phi).
\end{equation}
Here the $Y^m_n$ are spherical harmonics and the $A_{mn}$ and $B_{mn}$ are
constant expansion coefficients. The $\psi^{\pm}_{mn}$ diverge at
$\eta\rightarrow \pm \infty$, i.e. $x=y=0, z = \pm a$. Because the
potential in region I is regular, only the $\psi^{+}_{mn}$ can
contribute to the expansion. Likewise the expansions of the
cavity and reaction field inside the cavity only contain
  $\psi^{-}_{mn}$ terms. However, the field produced by the
dipole itself contains contributions from the divergent
$\psi^{+}_{mn}$ (see Appendix~\ref{app:b}). This is due to
the singularity of the potential at the centre of the sphere. We
therefore expand the potential in
each of the regions in the following series:
\begin{align}
  \label{eq:25}
  \psi_I &= \sum_{n=0}^\infty \sum_{m=-n}^n  A_{mn} \psi_{mn}^{+}\\ 
  \psi_{II} &= \sum_{n=0}^\infty \sum_{m=-n}^n \left[ B_{mn} \psi_{mn}^{-} +
    C_{mn} \psi_{mn}^{+}\right]\\
  \psi_{III} &= \sum_{n=0}^\infty \sum_{m=-n}^n \left[ D_{mn} \psi_{mn}^{-} +
    E_{mn} \psi_{mn}^{+}\right],
\end{align}
bearing in mind that the $E_{mn}$ are the known expansion coefficients
of the dipolar potential given by equations~\eqref{eq:43} and ~\eqref{eq:44}. When calculating the reaction field, the $E_{mn}$ vanish.
In this case the expansion coefficients of the external field, given
by equations \eqref{eq:45} and \eqref{eq:46}, have to be included in
the $B_{mn}$.
The expansion coefficients are determined by the boundary conditions.
Condition \eqref{eq:20}, i.e. continuity of the electric potential,
directly leads to
\begin{align}
  \label{eq:26}
  \sum_{n=0}^\infty \sum_{m=-n}^n A_{mn} \psi_{mn}^{+}\At{\eta=0} &= \sum_{n=0}^\infty \sum_{m=-n}^n \left[B_{mn} \psi_{mn}^{-} +
    C_{mn} \psi_{mn}^{+}\right]\At{\eta=0}\\
  \sum_{n=0}^\infty \sum_{m=-n}^n \left[B_{mn} \psi_{mn}^{-} +
    C_{mn} \psi_{mn}^{+}\right]\At{\eta=\eta_0} &= \sum_{n=0}^\infty \sum_{m=-n}^n \left[D_{mn} \psi_{mn}^{-} + E_{mn} \psi_{mn}^{+}\right] \At{\eta=\eta_0} 
\end{align}
for all $\vartheta$ and $\phi$. Multiplying these equations with
$Y^{M*}_N(\vartheta,\phi)/\sqrt{\cosh\eta-\cos\vartheta}$ and
integrating over $\sin\vartheta d\vartheta d\phi$, one is able to separate the
terms in the sums, with the result:
\begin{align}
  \label{eq:27}
  A_{MN} &=  B_{MN} + C_{MN}\\
  B_{MN} e^{-(N+1/2)\eta_0} + C_{MN} e^{(N+1/2)\eta_0} &=  D_{MN}
  e^{-(N+1/2)\eta_0} + E_{MN} e^{(N+1/2)\eta_0}. \label{eq:28}
\end{align}
The second condition, i.e. continuity of the normal component of the
displacement field, eq.~\eqref{eq:21}, leads to the equations
\begin{gather}
  \label{eq:29}
  \epsilon_1 \sum_{n=0}^\infty \sum_{m=-n}^n  A_{mn} \Diff{\psi_{mn}^{+}}{\eta}\At{\eta=0} =
    \epsilon_2 \sum_{n=0}^\infty \sum_{m=-n}^n \left[B_{mn} \Diff{\psi_{mn}^{-}}{\eta} +
    C_{mn} \Diff{\psi_{mn}^{+}}{\eta}\right]\At{\eta=0}\\
  \epsilon_2 \sum_{n=0}^\infty \sum_{m=-n}^n \left[B_{mn} \Diff{\psi_{mn}^{-}}{\eta} +
   C_{mn} \Diff{\psi_{mn}^{+}}{\eta}\right]\At{\eta=\eta_0} = \sum_{n=0}^\infty \sum_{m=-n}^n \left[D_{mn} \Diff{\psi_{mn}^{-}}{\eta} + E_{mn} \Diff{\psi_{mn}^{+}}{\eta}\right] \At{\eta=\eta_0} \label{eq:30}
\end{gather}
Projecting once more the sums, we find
\begin{equation}
  \label{eq:31}
  \epsilon_1 A_{MN} = \epsilon_2 (C_{MN}-B_{MN}).
\end{equation}
Equations \eqref{eq:27}, \eqref{eq:28} and \eqref{eq:31} can be used to express $A_{mn}$, $B_{mn}$, and $C_{mn}$ in term of the unknown $D_{mn}$.
\begin{align}
  \label{eq:32}
  A_{mn} &= \frac{2\epsilon_2}{\epsilon_2+\epsilon_1} \xi_n (D_{mn}+E_{mn} e^{(2n+1)\eta_0})\\
  B_{mn} &= \frac{\epsilon_2-\epsilon_1}{\epsilon_2+\epsilon_1} \xi_n
  (D_{mn}+E_{mn} e^{(2n+1)\eta_0})\\
  C_{mn} &= \xi_n (D_{mn}+E_{mn} e^{(2n+1)\eta_0})\\
\intertext{with}
\frac{1}{\xi_n} &= \frac{\epsilon_2-\epsilon_1}{\epsilon_2+\epsilon_1} + e^{(2n+1)\eta_0}.
\end{align}
Equation \eqref{eq:30} now reduces to
\begin{multline}
  \label{eq:33}
  \sum_{n=0}^\infty\sum_{m=-n}^{n} \biggl\{\left[\epsilon_2
  B_{mn}-D_{mn}\right]e^{-(n+1/2)\eta_0}\left[\frac{e^{-\eta_0}}{2}
  + n \cosh\eta_0\right]\biggr.\\
+ \left[E_{mn}-\epsilon_2  C_{mn}\right]
  e^{(n+1/2)\eta_0}\left[\frac{e^{\eta_0}}{2}+n\cosh\eta_0\right]-\left[(\epsilon_2B_{mn}-D_{mn})e^{-(n+1/2)\eta_0}\right.\\
 \biggl. \left. + (E_{mn}-\epsilon_2
  C_{mn})e^{(n+1/2)\eta_0}\right](n+1/2)\cos\vartheta \biggr\}Y^m_n(\vartheta,\phi) = 0.
\end{multline}
We multiply this equation with $Y^{M*}_N(\vartheta,\phi)$ and integrate over $\sin\vartheta
d\vartheta d\phi$. Using the product rule for spherical harmonics (see
eq. (A.26) of \cite{gray84}), we find
\begin{multline}
  \label{eq:34}
   \left[\epsilon_2
  B_{MN}-D_{MN}\right]e^{-(N+1/2)\eta_0}\left[\frac{e^{-\eta_0}}{2}
  + N \cosh\eta_0\right] \\
+ \left[E_{MN}-\epsilon_2  C_{MN}\right]
  e^{(N+1/2)\eta_0}\left[\frac{e^{\eta_0}}{2}+N\cosh\eta_0\right]=\\
\sum_{n=0}^\infty \sum_{m=-n}^n \left[(\epsilon_2 B_{mn}-D_{mn})e^{-(n+1/2)\eta_0}
   + (E_{mn}-\epsilon_2
  C_{mn})e^{(n+1/2)\eta_0}\right]\\
\times(n+1/2)\frac{2n+1}{2N+1}C(n1N;000) C(n1N;m0M),
\end{multline}
where  $C(n1N;m0M)$ are Clebsch-Gordan
coefficients. All terms with $m\neq M$ vanish, as well as all terms
with $n\neq N, N \pm 1$. The non-zero coefficients are
\begin{align}
  \label{eq:35}
  C(n1(n+1);m0m) &= \sqrt{\frac{(n-m+1)(n+m+1)}{(2n+1)(n+1)}},\\
  C(n1n;m0m) &= \frac{m}{\sqrt{n(n+1)}},\\
  C(n1(n-1);m0m) &= -\sqrt{\frac{(n-m)(n+m)}{n(2n+1)}}.
\end{align}
Equation \eqref{eq:34} therefore reduces to a three term recursion
coupling the $D_{mn}$, $D_{m(n-1)}$, and $D_{m(n+1)}$. When only a
finite number of terms are taken into account, the resulting
tridiagonal linear equation can be solved with standard methods. From
the $D_{mn}$ the electric field inside the cavity can be obtained
using equation \eqref{eq:41}. The series
$\sum_{n=0}^{\infty}\sum_{m=-n}^n D_{mn}
\psi_{mn}^{-}(\eta,\vartheta,\phi)$ converges very quickly at the
centre of the sphere, so that a truncation after 20 terms leads to
an accurate value for the field.

\begin{figure}[htbp]
  \centering
  \psfrag{epspar}{$\epsilon_2^{\|}$}
  \psfrag{epsperp}{$\epsilon_2^{\bot}$}
  \psfrag{epsperpasympt}{$\epsilon_2^{\bot}$ (asymptotic)}
  \psfrag{epsparasympt}{$\epsilon_2^{\|}$ (asymptotic)}
  \psfrag{deltaeps}{\LARGE $\epsilon_2(z)-\epsilon_2(\infty)$}
\includegraphics[angle=-90,width=\textwidth]{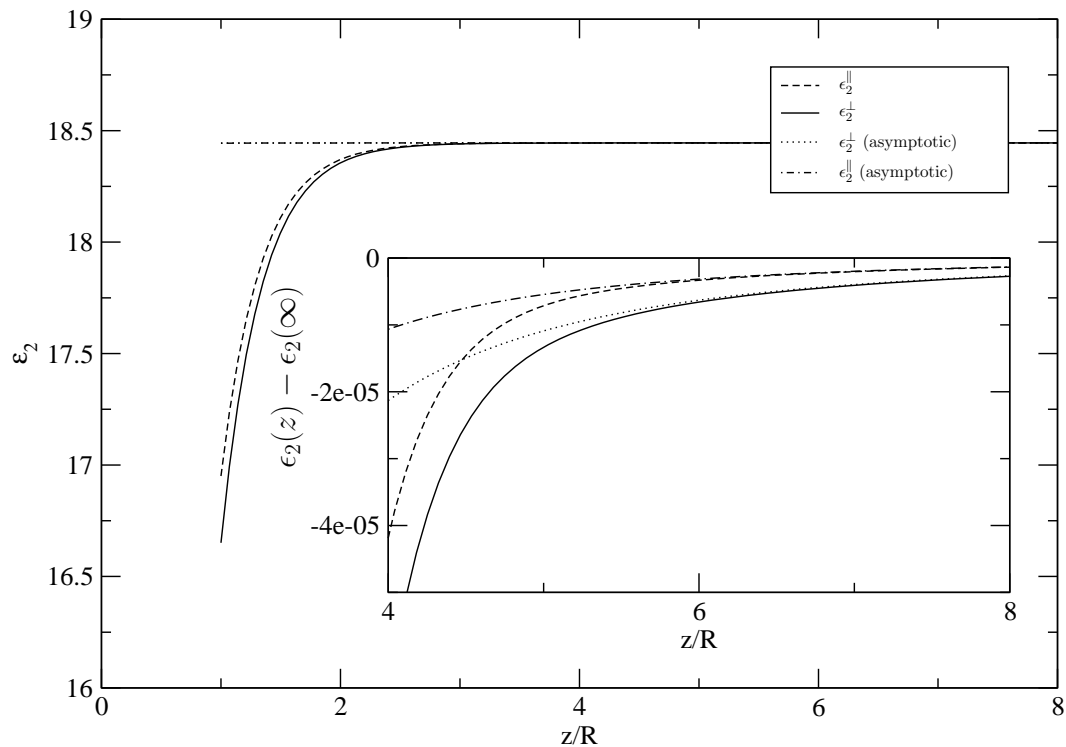}
  \caption{ Dielectric permittivity profile as calculated self
    consistently. Dipole moment and density of the fluid are the
    values measured for water (see caption Fig. \ref{fig:2}). The
    material below the interface is vacuum. The inset
    shows that the dielectric permittivities approach the values
    predicted by the asymptotic expansion \eqref{eq:13} very slowly.}
  \label{fig:4}
\end{figure}

After having calculated the reaction and cavity fields using a given
dielectric permittivity $\epsilon_2$, equations \eqref{eq:8} and
\eqref{eq:9} are employed to give new estimates for the permittivity
tensor. A locally self-consistent permittivity can thus be found by an
iterative procedure: We start by setting $\Tens{\epsilon_2}$ to the
isotropic bulk permittivity as determined by the bulk Onsager
relation~\eqref{eq:2}. We then solve the electrostatic equations as
outlined above, leading to $\Tens{\chi}_2^{\text{ext}}$.
Equations~\eqref{eq:8}, \eqref{eq:9} are then used to obtain a new
prediction of $\Tens{\epsilon}_2(z)$.  The new isotropic permittivity
$\overline{\epsilon_2} = \text{Tr}(\Tens{\epsilon}_2(z))/3$ is then
used as a starting value of the next iteration. In practice this
scheme converges very quickly. The resulting permittivity profile is
shown in figure \ref{fig:4}.

\section{Discussion}
\label{sec:discussion}

By generalising Onsager's model for the dielectric permittivity in
bulk fluids we were able to predict a permittivity profile of a
dipolar fluid near a wall. A number of approximations have been made;
some of which are of fundamental nature, and some are made to render
the calculations more tractable. A fundamental weakness of the Onsager
approach is the complete neglect of orientational correlations between
neighbouring molecules. This leads in the bulk to a significant
underestimation of the permittivity. In the vicinity of a dielectric
interface the fluid becomes birefringent. In our calculations we have
neglected the anisotropy of the permittivity, to simplify the
calculations. It is possible to take the full tensorial nature of the
permittivity into account, which would lead to a five term recurrence
formula instead of the three term recursion in equation \eqref{eq:34}.
However, the results (figure \ref{fig:4}) show that the fluid stays nearly isotropic except
in the immediate vicinity (i.e. a few molecular distances) of the
interface. Therefore the increased effort required to treat the
anisotropy exactly does not seem to be worthwhile. In order to derive
a semi-analytical expression for the permittivity, an inconsistency has
to be allowed for in treating the model. For each position of the
cavity the permittivity of the whole fluid is assumed to be constant.
The calculated permittivities, on the other hand, depend on the
distance of the cavity from the interface, leading to a permittivity
profile. This inconsistency can be overcome by solving the three
dimensional Laplace equation numerically starting from a given
permittivity profile. Again, the results of our model show bulk
behaviour nearly everywhere except in the immediate vicinity of the
interface, so that the complicated fully consistent solution does not
seem to be worthwhile. Far away from the wall we were able to derive a
first order correction to the bulk behaviour. We are presently
investigating how to incorporate the orientational correlations
between neighbouring dipoles into the theory.

\begin{acknowledgments}
RF is grateful to the Oppenheimer Fund for financial support. VB
acknowledges the support of the Royal Society and of the Swiss
National Science foundation.
\end{acknowledgments}


\begin{appendix}  

\section{Exact expression for the effective dipole moment}
\label{app:a}
Within a microscopic description of the dielectric medium modelled as
a classical dipolar fluid \cite{hansen86}, we can obtain an exact
expression for the effective dipole moment $\Vect{\mu}^{\text{eff}}$
of a given molecule inside the fluid. We assume the non polarizable molecules to have
a permanent dipole moment $\Vmu$, and we denote the
position and the orientation of the $i$th molecule by
$i=(\Ver_i,\Vmu_i)$, $i=1,...,N$. The molecules interact via 
of the dipolar pair potential
\begin{equation}
  v\Dip(1,2) = (\Vmu_1\cdot\Grad_1)(\Vmu_2\
  \cdot\Grad_2) \frac{1}{\Ds |\Ver_1-\Ver_2|}, \qquad |\Ver_1-\Ver_2|
  > 0,\label{eq:38}
\end{equation}
and a short-ranged interaction $v\Sr(1,2)$. The short-ranged interaction (e.g. a Lennard-Jones potential) is quite arbitrary, except for the fact that it must decay faster than $|\Ver_1-\Ver_2|^{-3}$ at large separations (it may thus include higher order multipole interactions between the molecules).

The dielectric constant $\epsilon$ of this fluid, at inverse
temperature $\beta$ and number density $\rho=N/V$, is given by the
Kirkwood formula \cite{kirkwood39}
\begin{equation}
  \label{Kirk}
  \frac{(\epsilon-1)(2\epsilon+1)}{9 \epsilon} = y \, g_{\text{{\tiny K}}},
  \qquad \quad y=4\pi\beta\rho \mu^2/9,
\end{equation}
where the factor $g_{\text{{\tiny K}}} = 1+\rho\int\Romd^3\Ver_1
\int\Romd\Omega_1\int\Romd\Omega_2\, (\hat{\Vmu}_1\cdot\hat{\Vmu}_2)
h(1,2)/(4\pi)^2$ is a measure of short-ranged angular correlations
between the molecules. The Onsager approximation to $\epsilon$,
equation \eqref{eq:2}, 
corresponds to neglecting these correlations, i.e. setting
$g_{\text{{\tiny K}}}=1$. We recall moreover that the Ursell function
$h(1,2)$ of the infinite system decays asymptotically like a screened
dipolar potential \cite{hansen86}.
\begin{equation}
  \label{h, r->oo}
  h(1,2) \sim \frac{(\epsilon-1)^2}{9 y^2 \epsilon} (-\beta v\Dip(1,2)),
  \qquad |\Ver_1-\Ver_2|\to\infty.
\end{equation}
To evaluate $\Vect{\mu}^{\text{eff}}$, we introduce
$\rho(1|2)=(\rho/4\pi)(1+h(1,2))$, the density of molecules at
$\Ver_1$ with orientation $\Vmu_1$ when there is a dipole $\Vmu_2$ at
$\Ver_2$. By definition, the average polarization of the fluid around
the fixed molecular dipole $\Vmu_2$ is
\begin{equation}
  \label{P(r)}
  \Vp(\Ver_1) = \int\Romd\Omega_1\, \Vmu_1 \, \rho(1|2) =
  \frac{\rho}{4\pi} \int\Romd\Omega_1\, \Vmu_1 \,
  h(1,2).
\end{equation}
This polarization cloud carries a total dipole moment
$\Vmu^{\text{cloud}} = \int \, \Ver \, (-\Grad\cdot\Vp(\Ver))\,
\Romd^3\Ver$. According to \eqref{P(r)} and \eqref{h, r->oo},
$\Vp(\Ver)$ decays like $|\Ver|^{-3}$ at large distances and is
therefore at the borderline of integrability. The integral defining
$\Vmu^{\text{cloud}}$ is nevertheless convergent, thanks to the
harmonicity of the Coulomb potential: $-\Grad\cdot\Vp(\Ver)$ decays
rapidly since $\Lapl(1/|\Ver|)=0$ when $|\Ver|>0$. An integration by
parts gives
\begin{equation}
  \label{int pol}
  \Vmu^{\text{cloud}} = \lim_{V\to\infty} 
  \left[ \int_V \Vp(\Ver)\,\Romd^3 \Ver - \int_{\partial V} \Ver \,
    (\Vp(\Ver)\cdot \Romd\Vect{S}) \right]
\end{equation}
for any volume $V$. For a spherical volume of radius $R\to\infty$, we find using \eqref{P(r)} and \eqref{Kirk} that the volume integral of the polarization is $(g_{\text{{\tiny K}}}-1) \Vmu_2$. The surface integral can be evaluated using \eqref{h, r->oo}, with the result $\Vmu_2 \, 2 (\epsilon-1)^2/(9 y \epsilon)$. This shows that the effective dipole moment is given by the simple result
\begin{equation}
\label{p^eff}
  \Vmu^{\text{eff}} = \Vmu + \Vmu^{\text{cloud}} =
  \frac{\epsilon-1}{3y\epsilon} \, \Vmu.
\end{equation}


\section{Properties of the bispherical coordinates}
\label{app:b}
The bispherical coordinates $(\eta,\vartheta,\phi)$ are
given by \cite{morse}
\begin{align}
  \label{eq:39}
  x &= \frac{a \sin \vartheta \cos \phi}{\cosh \eta - \cos \vartheta}\\
  y &= \frac{a \sin \vartheta \sin \phi}{\cosh \eta - \cos \vartheta}\\
  z &= \frac{a \sinh \eta}{\cosh \eta - \cos \vartheta}.
\end{align}
The surfaces of constant $\eta = \eta_0$ are spheres of radius $R = a
/ |\sinh \eta_0|$ with centre at the Cartesian coordinates $x=0, y=0,
z = a \coth \eta_0$. The bispherical coordinates of the centre (where
the dipole is located) are $\eta_1 = 2 \eta_0$,
$\vartheta_1 = 0$, and $\phi_1 = 0$. 
 
Given the coordinate system \eqref{eq:39} one can work out the unit
vectors in the $\eta$, $\vartheta$ and $\phi$-direction as
\begin{gather}
  \label{eq:40}
  \Vect{e}_\eta = \frac{1}{\cosh \eta - \cos \vartheta}
  \begin{pmatrix}
    \sin \vartheta \sinh \eta \cos \phi\\
    \sin \vartheta \sinh \eta \sin \phi\\
    1-\cosh \eta \cos \vartheta
  \end{pmatrix},\\
  \Vect{e}_\vartheta = \frac{1}{\cosh \eta - \cos \vartheta}
  \begin{pmatrix}
    (\cosh \eta \cos \vartheta - 1) \cos \phi\\
    (\cosh \eta \cos \vartheta - 1) \sin \phi\\
    \sinh \eta \sin \vartheta    
  \end{pmatrix}, \quad
  \Vect{e}_\phi =
  \begin{pmatrix}
    -\sin \phi\\
    \cos \phi\\
    0
  \end{pmatrix}.\nonumber
\end{gather}
The gradient is
\begin{equation}
  \label{eq:41}
  \Vect{\nabla} \psi =  \frac{\cosh \eta -\cos \vartheta}{a} 
  \left[\Diff{\psi}{\eta} \Vect{e}_\eta + \Diff{\psi}{\vartheta}
  \Vect{e}_\vartheta + \frac{1}{\sin \vartheta} \Diff{\psi}{\phi}
  \Vect{e}_\phi \right].
\end{equation}
The Laplace equation $\nabla^2 \psi = 0$ is separable in the
coordinate system and its solutions are of the form given in eq.~\eqref{eq:23}
Moreover, the Green's function, i.e. the solution of the equation $\nabla^2
G(\Vect{r}) = - \delta({\Vect{r} - \Vect{r}_1})$, is
\begin{align}
  \label{eq:42}
  G(\eta,\vartheta,\phi) &= \frac{1}{a} \sqrt{\cosh \eta - \cos \vartheta}\sqrt{\cosh \eta_1 -
    \cos \vartheta_1} \cdot\\
  & \qquad \sum_{n=0}^\infty \sum_{m=-n}^{n} \frac{4\pi}{2n+1}
  Y^{m*}_n(\vartheta_1, \phi_1) Y^m_n(\vartheta, \phi)
  e^{-(n+1/2)|\eta-\eta_1|}\\
  &= \frac{1}{a} \sum_{n=0}^\infty \sum_{m=-n}^n \frac{4\pi}{2n+1}
    \psi^{-*}_{an}(\eta_1,\vartheta_1,\phi_1) \psi^{+}_{an}(\eta,\vartheta,\phi).
\end{align}
The last line of the expansion is only valid for $\eta_1 > \eta$. Thus
we can derive the electrostatic potential of an elementary charge located
at any point. From there we obtain the potential of a dipole located
at the centre of the sphere $\eta=\eta_0$.

The potential for a dipole pointing in $z$-direction is obtained by taking
the derivative of the Green's function 
\begin{multline}
  \label{eq:43}
  \frac{\psi_{\text{dip}}^{\bot}(\eta,\vartheta,\phi)}{\mu}=\frac{\partial G}{\partial z_1} = \frac{\cosh \eta_1 - \cos
    \vartheta_1}{a} \frac{\partial G}{\partial \eta}\\
  = \frac{\sqrt{\cosh \eta_1 - 1}}{a^2}\sum_{n=0}^\infty 
  \sqrt{\frac{4\pi}{2n+1}}e^{- (n+1/2)\eta_1}
  \left[\frac{\sinh \eta_1}{2} - (n+1/2) (\cosh \eta_1-1)
    \right] \psi^{+}_{0n}(\eta,\vartheta,\phi).
\end{multline}
Similarly we find for a horizontal dipole in the $x$-direction
\begin{multline}
  \label{eq:44}
  \frac{\psi_{\text{dip}}^{\|}(\eta,\vartheta,\phi)}{\mu} = \frac{\partial G}{\partial x_1} = \frac{\cosh \eta_1 - \cos \vartheta_1}{a} \frac{\partial G}{\partial \vartheta}\\
  = \frac{(\cosh \eta_1 -1)^{3/2}}{a^2}\sum_{n=1}^\infty \frac{4\pi}{2n+1}
  e^{- (n +1/2) \eta}\sqrt{n(n+1)}
  [\psi^{+}_{1n}(\eta,\vartheta,\phi)-\psi^{+}_{-1n}(\eta,\vartheta,\phi)].
\end{multline}
From the generating function of the Legendre Polynomials we can find
the expansion of the vertical external electric field with amplitude
$E^{\text{ext}}$ valid in the region $\vartheta>0$
\begin{equation}
  \label{eq:45}
  \psi_{\text{ext}}^{\bot}(\eta,\vartheta,\phi) = -
  \sqrt{8\pi} a E^{\text{ext}}_{\bot}\sum_{n=0}^{\infty} \sqrt{2n+1} \psi_{0n}^{-}(\eta,\vartheta,\phi).
\end{equation}
For the horizontal external field in $x$-direction we obtain
\begin{equation}
  \label{eq:46}
 \psi_{\text{ext}}^{\|}(\eta,\vartheta,\phi) = 2 \sqrt{8\pi}
  a E^{\text{ext}}_{\|}\sum_{n=0}^\infty \sqrt{\frac{n(n+1)}{2n+1}}\left(\psi^{-}_{1n} - \psi^{-}_{-1n}\right).
\end{equation}

\end{appendix}


\end{document}